\documentclass[journal]{IEEEtran}
\usepackage{multicol}
\usepackage{multirow}
\usepackage{amsmath}
\usepackage{graphicx}
\usepackage{bm}
\usepackage{caption}
\usepackage{float}
\usepackage{algorithm}
\usepackage{algorithmicx}
\usepackage{algpseudocode}
\usepackage{cite}
\usepackage{url}
\usepackage[justification=centering]{caption}
\usepackage{color}
\usepackage{subfigure}
\usepackage{tabularx,colortbl}
\usepackage{makecell}
\usepackage{epstopdf}
\usepackage{mathrsfs}
\usepackage{amsmath, amsthm, amssymb}
\definecolor{blue1}{rgb}{0.00,0.50,1.00}
\pdfoutput=1
\begin{document}
\title{Robust Clock Synchronization via Low
Rank Approximation in Wireless Networks}

\author{Osama Elnahas,~\IEEEmembership{Member IEEE},
        Zhi Quan,~\IEEEmembership{Senior Member IEEE}

\thanks{Osama Elnahas and Z. Quan are with the College of Electronics and Information Engineering, Shenzhen University, Shenzhen 518060, China and Faculty of Information Technology, China. (Emails: {osama, zquan}@szu.edu.cn).}
\thanks{Osama Elnahas is also with the Department of Electrical Engineering, Faculty of Engineering, Assiut University, Egypt.}
}
\maketitle
\begin{abstract}

Clock synchronization has become a key design
objective in wireless networks for its essential
importance in many applications. However, as the wireless link
is prone to random network delays due to unreliable channel conditions, it is in general difficult to achieve accurate clock synchronization among wireless nodes.
This letter proposes robust clock
synchronization algorithms based on low rank matrix
approximation, which are able to correct timestamps in the
presence of random network delays. We design a low rank
approximation based maximum likelihood estimator (MLE) to
jointly estimate the clock offset and clock skew under
the two-way message exchange mechanism assuming Gaussian delay distribution.
By formulating the timestamp
correction problem into a low rank approximation problem,
we can solve the problem in the singular value
decomposition (SVD) domain and also via nuclear norm minimization.
Numerical results show that the proposed schemes can correct noisy timestamps and thus achieve more robust
synchronization performance than the MLE.

\end{abstract}

\begin{IEEEkeywords}
Clock synchronization, wireless
networks, low rank matrix approximation, maximum likelihood
estimator.
\end{IEEEkeywords}
\IEEEpeerreviewmaketitle
\section{Introduction}

Clock synchronization has emerged as a fundamental
requirement of many wireless applications to maintain, data fusion, coordination, power management and other common operations~\cite{Wu:survey2011}. For example, the coordination between wireless
nodes for power saving sleep/wake up schedules is based
on stable time agreement between all nodes. In sensor networks, clock synchronization is important to collect data from a physical environment and
then tag the data with the correct time of its occurrence. Clock synchronization is also essential for time related
transmission scheduling such as time division multiple access
(TDMA)~\cite{Wu:survey2011}.
In addition, many applications such as target tracking, localization, and
industrial control require highly precise clock synchronization among wireless nodes~\cite{Wu:survey2011}.

Every wireless node in the network has its own local clock. However, even if all clocks in the network are initially set accurately to a common time, they may drift away from each other over time. This is due to many factors such as oscillator imperfections and other environmental variations like temperature variation, and even differ with aging of the
clocks~\cite{Wu:survey2011}. Hence, it is essential to achieve clock synchronization between different nodes.

In recent years, many clock synchronization approaches have been developed for wireless networks such as Precision Time Protocol (PTP), network time protocol
(NTP), and global position system (GPS) to maintain time agreement between all nodes in
the network~\cite{Wu:survey2011}. For most applications, GPS is not appropriate as the wireless nodes have to be energy efficient,
low cost and may be designed for indoor usage. { Also, the
traditional NTP and PTP synchronization protocols that are
used in wired networks has been found to be unsuitable in
wireless networks due to random network delays, energy consumption, and communication overheads~\cite{ntpptp}.}
As a consequence, obtaining more simple, energy-efficient,
and accurate clock synchronization algorithms for wireless
networks is an important design objective.

The process of clock synchronization between any two wireless nodes is generally
achieved by timestamp packets exchanges. The collected timestamps are utilized to estimate the relative clock parameters,
i.e., the clock time phase offset (clock offset) and the clock
frequency offset (clock skew), and then adjust the clocks to
the common reference time~\cite{leng2010}. There have been substantial research efforts in clock synchronization
that correct only the clock offset without compensated
clock skew~\cite{shao2017}. However, adjusting only clock offset
requires a high re-synchronization rate to avoid frequent offset
change.
Hence, the joint correction of clock skew and offset will keep synchronization for a longer period, and therefore, lead to
energy saving and reduced communication overheads in
the network~\cite{leng2010}.

{One of the main challenges of clock synchronization over wireless networks is the random transmission delays of the
timestamp packets, which will make the clock parameters estimation process more complicated.
The network transmission delays can be divided into two main portions, the fixed delay and the random delay. In the literature, many random delays distributions have been assumed to be Gaussian, exponential, Weibull, and Gamma models. {The Gaussian and exponential distributions are commonly used delay distributions~\cite{Wu:survey2011},~\cite{E:serpedin2007}.}}

{In~\cite{E:serpedin2010}, the authors jointly estimated the fixed delay and the clock offset under the assumption of exponential random delays. The study in~\cite{E:serpedin2007} derived the maximum likelihood estimator (MLE) to jointly estimate the clock offset and skew  based on the assumption of Gaussian random delays and known fixed delay. This work was extended in~\cite{leng2010} to consider the case of exponential random delays.}

{Most of the existing works in the literature derive the MLE of the clock parameters using the collected noisy timestamps directly. However, noisy timestamps can affect the estimation process and thus deteriorate the
clock synchronization accuracy. In~\cite{E:serpedin2007}, it is shown that the Cramer-Rao lower bound (CRLB) is inversely proportional to the number of synchronization rounds. Thus, ML estimators require extra rounds in order to
improve the synchronization accuracy at the cost of communication overheads and energy consumption.
Hence, an important problem arises is how to design alternative clock synchronization scheme to denoise the collected set of timestamps.
Motivated by the above mentioned challenges, this letter proposes a robust timestamp correction scheme
for clock synchronization between two wireless nodes, which
can significantly denoise the collected timestamps with improved estimation accuracy.}

{The proposed algorithm has two complementary stages. In the first stage, timestamps correction
is formulated as a recovery of low-rank matrix from noisy timesatmps matrix.
Maximum likelihood estimator (MLE) is derived in the second stage to jointly estimate the
clock offset and skew from the denoised timestamps assuming Gaussian random delays
and unknown fixed delay.
We propose two timestamp
correction schemes based on low rank approximation, in
which the timestamp correction problem is formulated as a
low rank matrix approximation problem. We show that this
problem can be solved in the singular value decomposition
(SVD) domain and also by minimizing the nuclear norm of the matrix.
The simulation results show that the proposed algorithm can correct the collected set of
noisy timestamps and thus achieve more accurate estimation of clock parameters.}
\section{Clock Synchronization Model}

Consider clock synchronization between a reference Node $A$ with a reference time $T^{(A)}$ and another Node $B$ with time $T^{(B)}$ in a wireless network. Thus, at any particular time instant, the
relationship between the two clocks can be represented as
\begin{equation}
\label{eqn_clcokrelation}
T^{(A)}=\alpha T^{(B)}+\beta,
\end{equation}
{where $\alpha$ and $\beta$ represent the clock skew, i.e., frequency difference and clock offset, i.e., phase difference of Node $B$ with respect to Node $A$, respectively~\cite{E:serpedin2007}. If the two clocks are perfectly synchronized, then $\alpha=1$ and $\beta=0$. Due to the imperfections of the clock oscillator and environmental conditions, such as humidity and temperature the clock skew and offset vary over time.
The clock skew can be defined as the rate of clock variation and it varies with order of $10^{-6}$ for a typical
crystal-quartz oscillator~\cite{Wu:survey2011}.
The synchronization of Node $B$ to the reference Node $A$ requires the estimation of the unknown deterministic clock parameters $\alpha$ and $\beta$.}

\begin{figure}[!t]
  \centering
  \includegraphics[width=3.5in, height=1.65in]{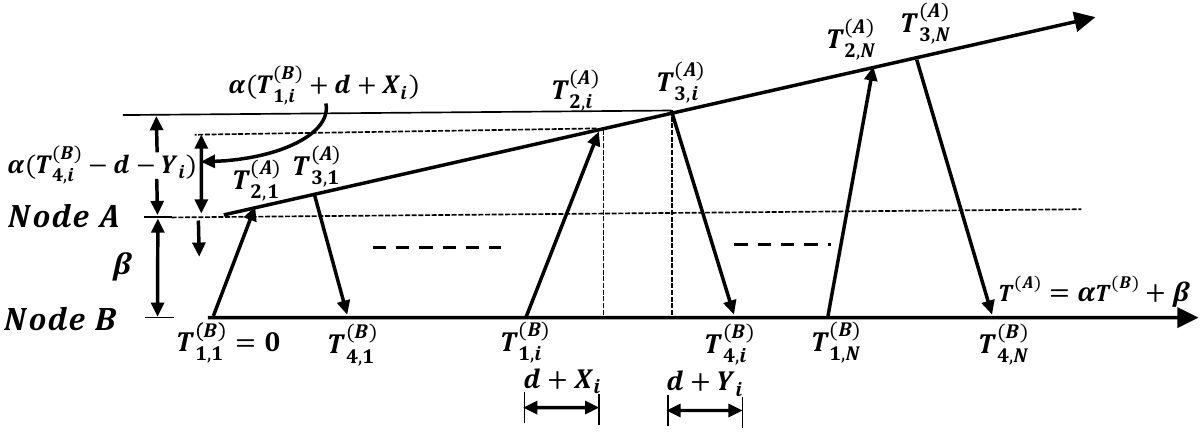}
  \caption{Two-way message exchange between Node A and Node B.}
  \label{two_way_exchange}
\end{figure}

Assume that the two nodes $A$ and $B$ are in the transmission range of each other. We consider {{the classical}} two-way message exchange mechanism for clock synchronization between
Node A and Node B.
In the two-way message exchange, two nodes $A$ and $B$
exchange N timestamp packets to perform the clock synchronization as shown in Fig.~\ref{two_way_exchange}.
In order to keep synchronized, the two nodes
need to periodically exchange these timestamp packets at a
specific synchronization rate.

 In the $i$-th round of packet exchange, node B records its current timestamp $T^{(B)}_{1,i}$ and sends a synchronization request to node A containing the value of $T^{(B)}_{1,i}$. Then, Node A records the reception
time of this timestamp packet $T^{(A)}_{2,i}$ according to its own local clock. After a certain time, the reference Node $A$ replies at $T^{(A)}_{3,i}$ with a timestamp packet to Node B containing $T^{(A)}_{2,i}$ and $T^{(A)}_{3,i}$. Finally, Node $B$ records its clock time $T^{(A)}_{4,i}$ at the reception of the
timestamp packet from Node A.
Note that the timestamps $T^{(B)}_{1,i}$ and $T^{(B)}_{4,i}$ are recorded by the clock of Node $B$, while the timestamps $T^{(A)}_{2,i}$ and $T^{(A)}_{3,i}$ are recorded by Node $A$. At the end of each synchronization cycle of $N$ rounds, Node $B$ has a set of timestamps $\left \{  T^{(B)}_{1,i}, T^{(A)}_{2,i}, T^{(A)}_{3,i}, T^{(B)}_{4,i} \right \}_{i=1}^{N}$. The goal of clock synchronization is to estimate the clock skew $
\alpha$ and the clock offset $\beta$ using the collected set of timestamps.
\section{The proposed Low Rank Estimator}
In this section, we present the proposed low rank approximation based timestamp correction scheme.
 Our main goal is to develop a scheme allowing us to jointly estimate the clock offset and skew in the presence of random network delays.
 We first introduce the joint estimation of clock skew and offset, and then show the proposed timestamp correction scheme.
\subsection{Joint Estimation of Clock Skew and Offset}
As shown in Fig.~\ref{two_way_exchange}, the collected timestamps in Node $B$ at the $i$-th synchronization round are expressed by
\begin{equation} \label{timestampT2}
\begin{split}
T^{(A)}_{2,i}=\alpha T^{(B)}_{1,i}+\beta+\alpha (d+X_i), \\
\end{split}
\end{equation}
 \begin{equation} \label{timestampT3}
\begin{split}
T^{(A)}_{3,i}=\alpha T^{(B)}_{4,i}+\beta-\alpha (d+Y_i),
\end{split}
\end{equation}
where $d$ refers to the fixed portion of the packet delay from one node to another, comprising the transmission time, propagation time, and reception time, $X_i$ and $Y_i$ represent the random portion of the packet delay in the uplink and downlink, respectively, including access time, send time, and receive time~\cite{leng2010}. In this letter, we assume that the uplink and downlink are symmetric and undergo the same amount of fixed delay $d$.
{By considering the random portion of the packet delay is a variable due to numerous independent random processes, we can assume that $X_i$ and $Y_i$ are modeled as independent and identical distributed (i.i.d.) Gaussian random
variables with zero mean and variance $\sigma_\text{n}^2$. This assumption has been experimentally tested in~\cite{gaussianmodel}.}
In the following, we derive the MLE of the clock offset and skew in the presence of normally distributed transmission delays. First, we can rewrite~\eqref{timestampT2} and~\eqref{timestampT3} as~\cite{E:serpedin2007}
\begin{align}
\label{substracttimestamps}
T^{(B)}_{1,k}=\frac{1}{\alpha}T^{(A)}_{2,k}-\frac{\beta}{\alpha}-d-G_k,
\end{align}
\begin{align}
\label{substracttimestamps}
-T^{(B)}_{4,k}=-\frac{1}{\alpha}T^{(A)}_{3,k}+\frac{\beta}{\alpha}-d-H_k.
\end{align}
Then, we stack all the $N$ synchronization rounds into a matrix form as
\begin{equation}
\label{stackingtimestamps}
\underbrace {\left[ \begin{array}{l}
T^{(B)}_{1,1}\\
{~~} \vdots \\
T^{(B)}_{1,N}\\
-T^{(B)}_{4,1}\\
{~~} \vdots \\
-T^{(B)}_{4,N}
\end{array} \right]}_{ \buildrel \Delta \over = {\textbf{T}_B}} = \underbrace {\left[ \begin{array}{l}
{T^{(A)}_{2,1}}~~~~ {-1}~~~ {-1}\\
~~\vdots ~~~~~~~~~ \vdots~~~~~~ \vdots\\
{T^{(A)}_{2,N}}~~~ {-1}~~~ {-1}\\
{-T^{(A)}_{3,1}}~~~~ {1}~~~ {-1}\\
~~\vdots ~~~~~~~~~ \vdots~~~~~~ \vdots\\
{-T^{(A)}_{3,N}}~~~ {1}~~~ {-1}
\end{array} \right]}_{ \buildrel \Delta \over = {\textbf{T}_A}} \underbrace {\left[ \begin{array}{l}
\psi_1 \\
\psi_2\\
\psi_3\\
\end{array} \right]}_{ \buildrel \Delta \over = {\pmb{\Psi}}}- \underbrace {\left[ \begin{array}{l}
{X_1} \\
{~} \vdots \\
{X_N} \\
{Y_1} \\
{~} \vdots \\
{Y_N}
\end{array} \right]}_{ \buildrel \Delta \over = \textbf{Z}},
\end{equation}
where $\psi_1=1/\alpha$, $\psi_2={\beta}/{\alpha}$, and $\psi_3=d$.

 We assume that the fixed portion of packet delay $d$ is unknown and the random delays $\left \{  X_k \right \}_{i=1}^{N}$ and $\left \{  Y_k \right \}_{i=1}^{N}$ are normally distributed with zero mean $\mu=0$ and variance $\sigma^2_\text{n}$, i.e., ${X_i}\sim\mathcal{N}(0,{\sigma_{\text n} ^2})$, ${Y_i}\sim\mathcal{N}(0,{\sigma_{\text n} ^2})$. Then,~{the likelihood function} for $(\textbf{T}_A, \textbf{T}_B, \pmb{\Psi})$, based on the collected set of timestamps $\left \{  T^{(B)}_{1,i}, T^{(A)}_{2,i}, T^{(A)}_{3,i}, T^{(B)}_{4,i} \right \}_{i=1}^{N}$ is given by~{{\cite{leng2010}}}
{\begin{equation}
\label{likelihoodfunction}
  \begin{aligned}
&f(\textbf{T}_A, \textbf{T}_B, \pmb{\Psi})
   &=(2 \pi \sigma^2_\text{n})^{-N}e^{{-\frac{{{\left \| \textbf{T}_B -\textbf{T}_A \pmb{\Psi} \right \|}^2}}{2 \sigma^2_\text{n}}}}.
\end{aligned}
\end{equation}}
where $\textbf{T}_A$, $\textbf{T}_B$, and $\pmb{\Psi}$ are defined in~\eqref{stackingtimestamps}.
 The log-likelihood function for $(\textbf{T}_A, \textbf{T}_B, \pmb{\Psi})$ can be written as
 {\begin{equation}
\label{likelihoodfunction2}
  \begin{aligned}
\text{ln} f(\textbf{T}_A, \textbf{T}_B, \pmb{\Psi})=N\text{ln} \frac{1}{2 \pi \sigma^2_\text{n}}-\frac{{\left \| \textbf{T}_B -\textbf{T}_A \pmb{\Psi} \right \|}^2}{2\sigma^2_\text{n}},
\end{aligned}
\end{equation}}
For a given set of timestamps, taking the derivative over the log-likelihood function defined in~\eqref{likelihoodfunction2} with respect
to $\pmb{\Psi}$, and setting the result to zero, the MLE of $\pmb{\Psi}$ is given by
 \begin{equation}
\label{mleofphi}
  \begin{aligned}
\hat{\pmb{\Psi}}=\left ( \textbf{T}_A^H\textbf{T}_A \right )^{-1}\textbf{T}_A^H \textbf{T}_B.
\end{aligned}
\end{equation}
Finally, we obtain the estimated clock skew $\hat \alpha=\hat{1/\pmb{\Psi}}[1]$, clock offset $\hat \beta=\hat{\pmb{\Psi}}[2]/\hat{\pmb{\Psi}}[1]$, and the fixed delay $\hat{\pmb{\Psi}}[3]$, where $w[j]$ denotes $j$-th element of vector $w$~{\cite{leng2010}}.
~{The CRLBs for the joint estimation of clock skew $\hat \alpha$ and clock offset $\hat \beta$ can
be derived by forming the Fisher information matrix for~(\ref{likelihoodfunction2}).
Accordingly, the CRLBs of the estimated clock offset and clock skew are given as~\cite{leng2010}
\begin{equation} \label{crlbskew}
\begin{split}
\text{CRLB}(\hat \alpha) = \frac{2N \sigma_{\text n}^2}{2N\mathcal{U}-\alpha^2 \mathcal{V}^2-\mathcal{W}^2},~~~~~
\end{split}
\end{equation}
and
\begin{equation} \label{crlboffset2}
\begin{split}
\text{CRLB}(\hat \beta) = \frac{\sigma_{\text n}^2 \alpha^2 (2N \mathcal{U}-\mathcal{V}^2)}{2N(2N\mathcal{U}-\alpha^2 \mathcal{V}^2-\mathcal{W}^2)},
\end{split}
\end{equation}
where $\mathcal{U}$, $\mathcal{V}$, and $\mathcal{W}$ are as follows~\cite{leng2010}
\begin{equation} \label{eqA1}
\begin{split}
\mathcal{U}=\frac{1}{\alpha^4}\sum_{i=1}^{N}\left [ \alpha^2(T_{1,i}+d)^2+\alpha^2 \sigma_{\text n}^2+(T_{3,i}-\beta)^2 \right ]
\end{split}
\end{equation}
 \begin{equation} \label{eqA2}
\begin{split}
\mathcal{V}=\frac{1}{\alpha^3}\sum_{i=1}^{N}\left [ \alpha(T_{1,i}+d)+(T_{3,i}-\beta) \right ]~~~~~~~~~~~~~~~
\end{split}
\end{equation}
\begin{equation} \label{eqA3}
\begin{split}
\mathcal{W}=\frac{1}{\alpha^2}\sum_{i=1}^{N}\left [ \alpha(T_{1,i}+d)-(T_{3,i}-\beta) \right ].~~~~~~~~~~~~~~
\end{split}
\end{equation}}
The proposed scheme forms a timestamp matrix ${\pmb G} $ to hold the collected set of timestamps $\left \{  T^{(B)}_{1,k}, T^{(A)}_{2,k}, T^{(A)}_{3,k}, T^{(B)}_{4,k} \right \}_{k=1}^{N}$ from the $N$ synchronization rounds. In the timestamp matrix ${\pmb G}$, a row corresponds to the synchronization round and a column corresponds to timestamps of the round.

As shown in Fig.~\ref{two_way_exchange}, Node $B$ sends timestamp packets in a constant interval between synchronization rounds ${\delta T^{(B)}}=T^{(B)}_{1,i}-T^{(B)}_{1,i-1}$. Therefore, the collected timestamps usually have strong correlation.
Denote ${\mathbf h}^T = [0, D, D+b, 2D+b]$ and ${\mathbf t} = [T^{(B)}_{1,1}, T^{(B)}_{1,2}, \ldots, T^{(B)}_{1,N}]^T$, where $b$ refers to the processing delay at Node $A$, and $D$ represents the transmission delay assuming all the delays are fixed.
Then it is easy to write the timestamp matrix as ${\mathbf G} = {\mathbf 1}_N{\mathbf h}^T + {\mathbf t}{\mathbf 1}_4 ^T$ where ${\mathbf 1}_N$ stands for an all-one vector of length $N$. Hence ${\mathbf G}$ is a rank-two matrix.
Thus, if we re-arrange the collected timestamps to a matrix, such a matrix become a noisy version
of a low-rank matrix. Thus, the problem of correcting timestamps is converted to
the problem of recovering a low-rank matrix from its noisy
version.

Suppose that the noise free timestamp matrix $\pmb G \in \mathbb{R}^{N\times L}$ has a rank $k$ and the noisy timestamp matrix $\pmb G_\text{n}$ can be defined as
\begin{equation} \label{noisymatrix}
\begin{split}
{\pmb G}_\text{n}={\pmb G}+{\pmb Z},
\end{split}
\end{equation}
where~{$L$ is the number of timestamps in each synchronization round, i.e., $L=4$}, and ~{$\pmb Z$} represents the additive noise caused by the random delays. Now our task is to estimate the matrix $\pmb G$ as accurately as possible from its
noisy version ${\pmb G}_\text{n}$. The proposed scheme obtains an estimate of the denoised matrix $\hat {\pmb G}$
by approximating the noisy matrix ${\pmb G}_\text{n}$ with another low rank matrix that matches all the received entries by solving the following optimization problem
\begin{equation} \label{joint MC recovery}
\begin{split}
\hspace{0.2cm}{\min} \hspace{0.2cm}&  \text{rank}(\hat{\pmb G}) \\
 \text{s.t.} \hspace{0.2cm} & \left \| \hat{\pmb G}-{\pmb G_\text{n}}\right \|_F <  \eta, \\
\end{split}
\end{equation}
where~{ $\hat{\pmb G}$ is an estimate of the denoised low rank matrix}, and rank($\hat{\pmb G}$) denotes the rank of the estimated matrix $ \hat{\pmb G}$.~{Besides, $\eta > 0 $ is regularization parameter controlling the tolerance error of the minimizer.}~{In the following subsections, we propose two methods to approximate the noisy matrix ${\pmb G}_\text{n}$ with another low rank matrix $\hat {\pmb G}$.} First, we utilize the energy distribution in the SVD domain
to estimate the low-rank approximation of the noisy timestamp matrix by only preserving a certain number of the largest singular values.
Moreover, a low rank matrix approximation (LRMA) model is applied to approximate the noise-free timestamp matrix by nuclear norm minimization.
\subsection{Denoising Via Low-rank Approximation in SVD Domain}
\label{svddenoising}
In the SVD domain, the noisy timestamp matrix $\pmb G_\text{n} \in \mathbb{R}^{N\times L}$ can be expressed as
\begin{equation} \label{svd}
\pmb G_\text{n}=\pmb U \pmb \Sigma \pmb V^T,
\end{equation}
where $\pmb U$ is an $N \times N$ orthogonal matrix, $\pmb V$ is a an $L\times L$ orthogonal matrix with $\pmb V^T$ its transpose, and $\pmb \Sigma$ is a $N \times L$ diagonal matrix with the diagonal entries represent the singular values of the matrix $\pmb G_\text{n}$.
If most of the energy is occupied in its first top $k$ singular values, the matrix is said to be low rank.
However, due to the noise matrix $\pmb Z$, the energy of delay noise will span all the singular values and the last {$p-k$} singular values of~{$\pmb G_\text{n}$} will not be zero, where~{$p$ is the total number of singular values}.

The proposed scheme gets the denoised low rank version of the matrix~{$\pmb G_\text{n}$} based on~\eqref{joint MC recovery}, where it keeps a certain number of largest singular values and
the small singular values are truncated to zero. This process is reasonable as the largest singular values carry the most energy of the matrix and the small ones correspond to noise. Thus, the low rank version $\hat{\pmb G}$ of the noisy matrix $\pmb G_\text{n}$ is given by
\begin{equation} \label{svd_solution}
 \hat{\pmb G}=\pmb U \pmb \Sigma_k \pmb V^T,
\end{equation}
where $\pmb \Sigma_k$ keeps only the first $k$ singular values of the diagonal matrix~{$\pmb \Sigma=\text{diag}(\sigma_1, \dots , \sigma_p)$} and set the other diagonal values to zero. Therefor, $\pmb \Sigma_k$ can be defined by
 \begin{equation} \label{svd_solution2}
 \pmb \Sigma_k=\text{diag}(\sigma_1, \dots, \sigma_k, 0, \dots, 0),\,
\end{equation}
where $\sigma_i\geq 0$ is the $i$-th singular value of the matrix $\pmb G_\text{n}$.~{The obtained result in~(\ref{svd_solution2}) is defined as the matrix approximation lemma or Eckart-Young-Mirsky theorem~\cite{eckart}.}
Then, the proposed scheme use the denoised low rank timestamp matrix to estimate the clock offset and clock skew based on~\eqref{mleofphi}.
\subsection{LRMA Denoising}
The proposed SVD based denoising scheme in~\ref{svddenoising} assumes that the noise concentrates on the smallest singular values and preserves only the largest singular values. This hard-threshold solution ignores that the noise energy also spreads among the largest singular values in the SVD domain, specially in case of high noise level.~{As an alternative, we propose a convex low rank matrix approximation that minimizes the nuclear norm of the low rank matrix.} This method
aims to transfer the noisy timestamp matrix $\pmb G_\text{n}$ into a low rank version with a more robust formulation.

As discussed above, the timestamp matrix $\pmb G$ with dimension $N \times L$ holds the {{collected set of timestamps}}.
The entries of matrix $\pmb G$ are perturbed by the random transmission delays, thus producing a noisy matrix $\pmb G_\text{n}$.
The computational complexity of the rank minimization problem in~(\ref{joint MC recovery}) is NP-hard.
Inspired by the convex relaxation alternative in~\cite{matrixcompletionwithnoise},
we introduce the convex matrix rank approximation to estimate $\hat {\pmb G}$ from the noisy timestamp matrix $\pmb G_\text{n}$ by minimizing the nuclear norm of the matrix. The proposed scheme is defined as
\begin{equation} \label{matrixcompletionconvex}
\begin{split}
{\min} \hspace{0.2cm}&  \left \| \hat {\pmb G} \right \|_* \\
 \text{s.t.} \hspace{0.2cm} & \left \| \hat{\pmb G}-{\pmb G_\text{n}}\right \|_F <  \eta,
\end{split}
\end{equation}
 where $\left \| \hat {\pmb G} \right \|_*$ is the nuclear norm of the matrix $\hat {\pmb G}$. It is defined as
 \begin{equation}
  \left \| \hat {\pmb G} \right \|_*=\sum_{i=1}^{\min\left \{ N,L \right \}}\sigma_i.
  \end{equation}
There are many matrix rank minimization algorithms to solve~(\ref{matrixcompletionconvex}) in an efficient way~\cite{spectralmcieee}.~{In our work, we use the efficient rank minimization algorithm proposed in~\cite{spectralmcieee}, based on
spectral methods followed by a local manifold optimization~{{due to}} its simplicity in implementation.}


\section{Numerical Results}

In this section, we numerically evaluate the performance of the two proposed low rank-based clock synchronization schemes.~{We set the involved clock parameters as follows: the clock skew and clock offset are chosen uniformly within the intervals $\alpha \in (0.99, 1.01)$ and $\beta \in [-10, 10]$, respectively~\cite{leng2011}. Besides, we choose the fixed delay as $d \in [1, 10]$ and the random delays $\left \{  X_k \right \}_{k=1}^{N}$ and $\left \{  Y_k \right \}_{k=1}^{N}$ are normally distributed with zero mean $\mu=0$ and variance $\sigma_{\text n}^2=1$, which are used frequently in the literature~\cite{leng2011}. For all the conducted simulations, the CRLB is calculated based on~(\ref{crlbskew}),~(\ref{crlboffset2}), and the estimated noise variance $\hat \sigma^2_\text{n}$ after LRMA denoising. In the SVD denoising, the rank $k$ is set to $k=2$.}
%
\begin{figure}[!t]
  \centering
  \includegraphics[width=3.5in, height=2.5in]{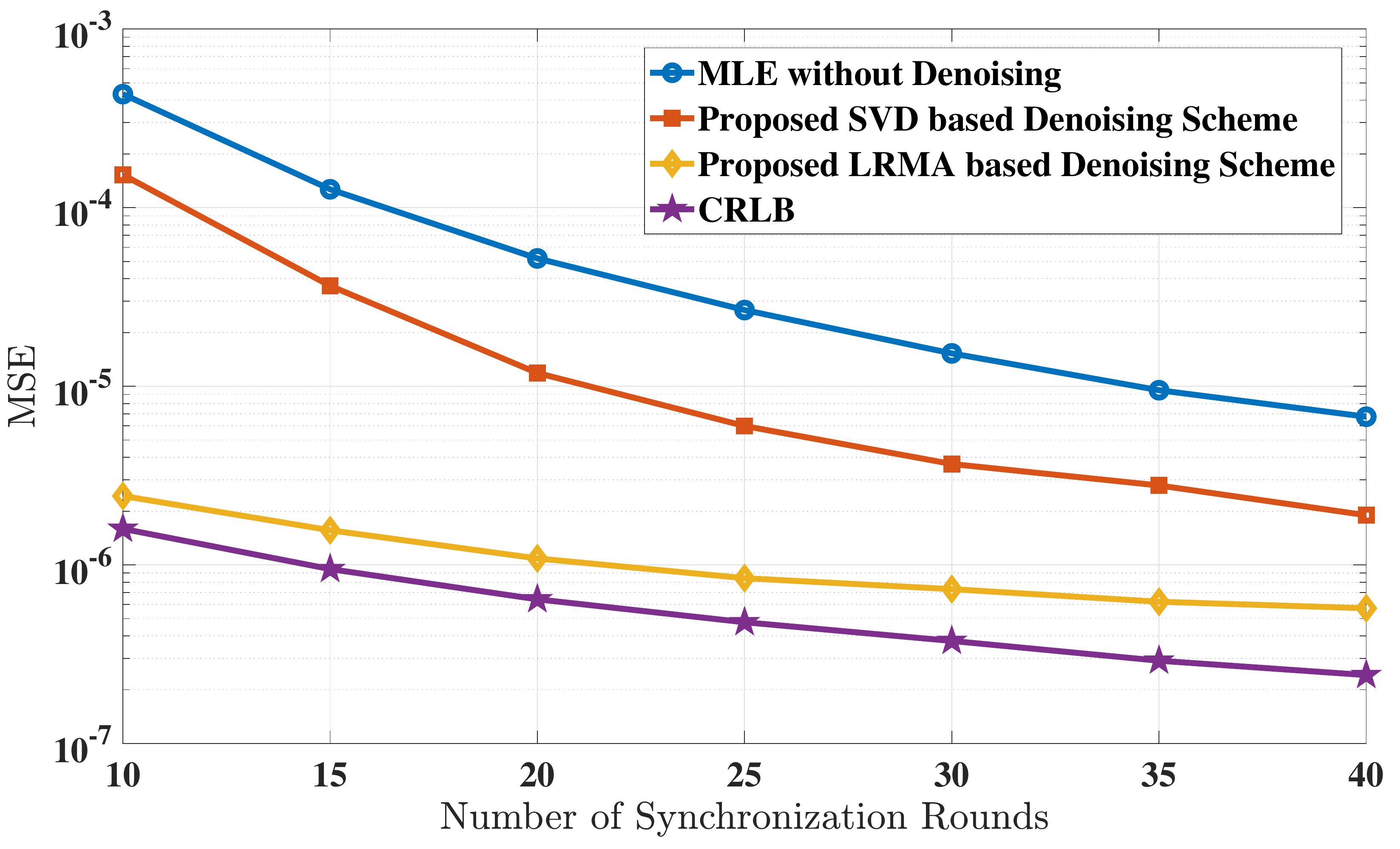}
  \caption{MSE of the estimated clock skew $\hat \alpha$ v.s. Number of synchronization
rounds.}
  \label{fig_itteration}
\end{figure}

{Fig.~\ref{fig_itteration} illustrates the mean squared error (MSE) of the estimated clock skew
versus the number of synchronization rounds using different estimators.}
{Obviously, for all schemes, the MSE decreases as the number of synchronization rounds $N$ becomes large.
From Fig.~\ref{fig_itteration}, It can be seen that
the proposed low rank-based schemes significantly improves the estimation accuracy compared to the MLE scheme.
The reason for this is that the proposed schemes corrected the collected set of timestamps, then estimates the clock parameters, while the MLE scheme uses the noisy timestamps directly.}

\begin{figure}[!t]
  \centering
  \includegraphics[width=3.5in, height=2.5in]{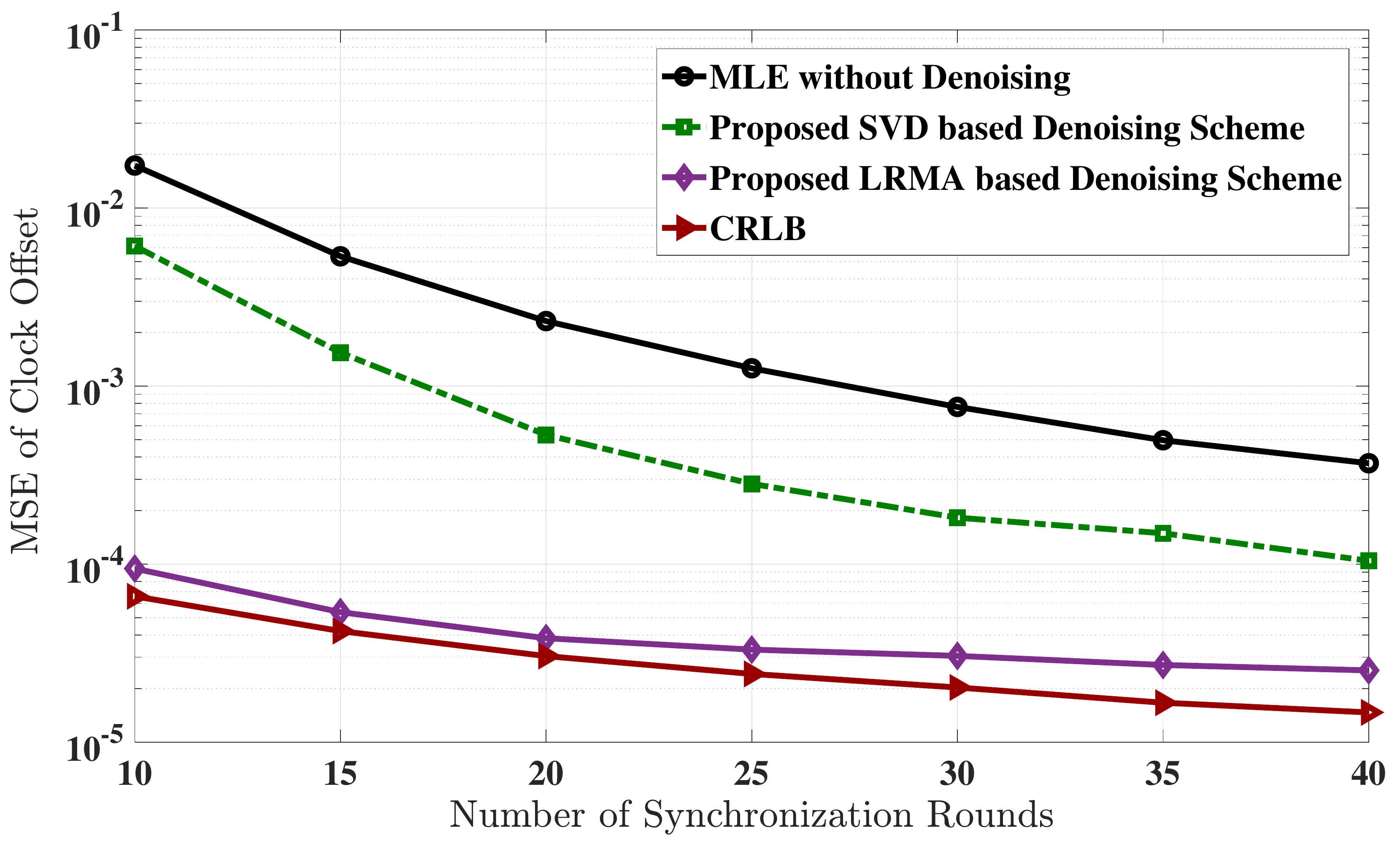}
  \caption{MSE of the estimated clock offset $\hat \beta$ v.s. Number of
synchronization rounds.}
  \label{fig_Error}
\end{figure}

{Fig.~\ref{fig_Error} presents MSE for estimation of the clock offset with the same conclusion as Fig.~\ref{fig_itteration}. Besides, it is shown from Fig.~\ref{fig_itteration} and~\ref{fig_Error} that the LRMA denoising scheme yields lower MSE for
all values of $N$ than the SVD based one.
This is due to the truncation in the SVD based scheme, which assumes that most of information energy concentrates on the first $k$ largest singular values and noise concentrates in the remaining ones. However, the noise may spread among the first $k$ singular values when high noise level exists. Note that the performance of the LRMA denoising scheme is close to the CRLB as shown in Fig.~\ref{fig_itteration} and~\ref{fig_Error}}.

{In addition, it is obvious from Fig.~\ref{fig_itteration} and~\ref{fig_Error} that the proposed LRMA scheme is robust with high estimation accuracy even with a reduced number of synchronization rounds. This means that the proposed LRMA scheme can achieve relatively high synchronization accuracy at the price of less communication overheads and energy consumption. Also, it is noticed that the proposed LRMA scheme almost has the same behavior even for different number of synchronization rounds.}

{As discussed in the previous sections, the proposed scheme has two stage, the timestaps correction stage and the clock parameters estimation stage. The computational complexity of the estimation stage is same as that of MLE scheme with order $\mathcal{O}(N)$. The timestamps correction stage includes SVD denoising and LRMA denoising with complexity $\mathcal{O}(16N)$ and $\mathcal{O}(4Log(4)Nr)$, respectively for the timestamps matrix $\pmb G \in \mathbb{R}^{N\times 4}$ with rank $r$~\cite{spectralmcieee}.
In fact, the proposed scheme has improved performance compared to MLE estimators at slightly higher computational complexity and reduced number of packets transmissions. As shown in Fig.~\ref{fig_itteration} and~\ref{fig_Error}, the MLE scheme requires additional overheads, i.e. packet exchanges for more accurate estimation.
In~\cite{Pottie}, it has been shown that the energy cost for transmitting $1$ kbit a distance of $100$ m is approximately equivalent to the energy needed to execute three million instructions. Consequently, the proposed scheme trades-off between the computational overhead and reduced communication energy without incurring any loss of parameters estimation accuracy.}
\section{Conclusion}

In this letter, we have presented low rank-based denoising
schemes for joint estimation of clock skew and offset based
on the two-way message exchange mechanism assuming Gaussian delay distribution. We
proposed two clock synchronization schemes, low rank
denoising in the SVD domain and the LRMA denoising. By
comparing with the MLE scheme, we find that both
schemes achieve better performance than the MLE. In
addition, among the two proposed schemes, the proposed
LRMA denoising scheme is superior to the SVD based
scheme, due to the truncation utilized by the SVD based
scheme. {The proposed scheme is built on the robust property of low rank approximation models that are robust to various types of noise. Therefore, the proposed approach can be generalized to other random delay distributions like exponential, Weibull, Gamma, or even unknown network distribution.}

\bibliographystyle{iEEEtran}

\end{document}